\documentclass[runningheads]{llncs}
\usepackage[T1]{fontenc}
\usepackage{graphicx}
\usepackage[linesnumbered,ruled,vlined]{algorithm2e}
\usepackage{amsmath}
\usepackage{booktabs}
\usepackage{syntax}
\usepackage{mathtools}
\usepackage{listings}

\usepackage[table]{xcolor}

\usepackage{colortbl}
\usepackage[english]{babel}
\usepackage{multirow}

\definecolor{javared}{rgb}{0.6,0,0} %
\definecolor{javagreen}{rgb}{0.25,0.5,0.35} %
\definecolor{javapurple}{rgb}{0.5,0,0.35} %
\definecolor{javadocblue}{rgb}{0.25,0.35,0.75} %
\definecolor{javagrey}{rgb}{0.46,0.45,0.48} %

\usepackage[most]{tcolorbox}

\newtcolorbox{resqbox}{
  enhanced,
  breakable,
  colback=green!5,
  colframe=green!40!black,
  boxrule=0.8pt,
  left=4pt,right=4pt,top=4pt,bottom=4pt,
}

\newcommand\resq[1]{%
\noindent\begin{resqbox}
#1
\end{resqbox}
}

\lstdefinestyle{Alg}{
  basicstyle=\ttfamily\footnotesize,
  breaklines=true,
  tabsize=2,
  mathescape,
  numbers=left,
  xleftmargin=2.5em,
  xrightmargin=0.5em,
  frame=tb,
  framexleftmargin=2em,
  emph={Algorithm,Input,Output,for,each,do,if,else,Function,while,let,be,repeat,until,return,times,and,or,break,in,then,},
  emphstyle={\textbf},
  escapechar=?,
  morecomment=[l][\color{javagreen}]{//},
  columns=flexible,
}

\lstdefinestyle{Spec}{
	language=Java, %
	keywordstyle=\color{javapurple}, %
	stringstyle=\color{javared},
	commentstyle=\color{javagreen},
	morecomment=[s][\color{javadocblue}]{/**}{*/},
	morecomment=[l][\color{javagrey}]{@},
	morecomment=[l][\color{javagrey}]{//@},
    morecomment=[l][\color{javagrey}]{/*@},
    morecomment=[l][\color{javagrey}]{*/},
	basicstyle=\ttfamily\footnotesize,
	breaklines=true,
	tabsize=2,
	frame=single,
	mathescape,
	numbers=left,
	xleftmargin=2.5em,
	xrightmargin=0.5em,
	frame=single,
	framexleftmargin=2em,
	morekeywords={,duration,simulationNodes,ms,Platform,CPU,memory,SimulationNode,platform,cloud,Cloud,IP,port,protocol,b,Simulator,username,password,quanta,step,Device,period,payload,speed,devices,locationIP, G, EdgeDevices,workload, inToOut, EdgeDevice,type, offsetRange, s, pubTopic, subTopic, }
	escapeinside={\%*}{*)}, %
    columns=flexible,
	escapechar=?,
}

\lstdefinestyle{myCustomMatlabStyle}{
  language=Matlab,
  numbers=left,
  stepnumber=1,
  numbersep=10pt,
  tabsize=4,
  showspaces=false,
  showstringspaces=false
}

\usepackage{seqsplit}
\usepackage{framed}
\definecolor{light-gray}{gray}{0.9}
\usepackage[framemethod=TikZ]{mdframed}
\mdfsetup{skipabove=5pt,skipbelow=3pt}
\usepackage{lipsum}
\mdfdefinestyle{MyFrame}{%
	linecolor=black,
	outerlinewidth=0.15pt,
	roundcorner=3pt,
	innertopmargin=2pt,
	innerbottommargin=2pt,
	innerrightmargin=4pt,
	innerleftmargin=4pt,
	backgroundcolor=light-gray}

\makeatletter
\let\oldthebibliography\thebibliography
\let\endoldthebibliography\endthebibliography
\renewenvironment{thebibliography}[1]{%
  \oldthebibliography{#1}%
  \fontsize{7.6pt}{9.6pt}\selectfont      %
}{%
  \endoldthebibliography
}
\makeatletter

\newcommand{\framework}{AutoSLO}

\begin{document}
\title{Genetic Programming for Self-Adaptive Auto-Scaling of Microservices}
\titlerunning{Self-Adaptive Auto-Scaling of Microservices}
\author{Jia Li\inst{1}\orcidID{0009-0000-8859-4541} \and Mehrdad Sabetzadeh\inst{1}\orcidID{2222--3333-4444-5555} \and Shiva Nejati\inst{1}\orcidID{1111-2222-3333-4444}}
\authorrunning{J. Li, M. Sabetzadeh, S. Nejati}
\institute{University of Ottawa, 800 King Edward Ave, Canada
\email{jli714,m.sabetzadeh,snejati@uottawa.ca}}
\maketitle              %
\begin{abstract}
Microservice architecture is widely adopted in modern systems, where auto-scaling is critical for satisfying service-level objectives (SLOs). However, determining optimal scaling for microservices is difficult, and reactive resource allocation often leads to costly over- or under-provisioning. We propose AutoSLO, a learning-based, self-adaptive scaling framework that dynamically adjusts microservice replicas to meet SLOs while minimizing resource usage. AutoSLO uses a continuous \linebreak monitoring-adaptation feedback loop and leverages genetic programming to learn and evolve scaling logic, enabling the deployed microservice system to proactively prevent SLO violations rather than repeatedly searching for one-off scaling actions. We evaluate AutoSLO on two case-study systems -- an online shopping platform and a chatbot based on large language models -- and show that this framework substantially reduces resource usage while maintaining a low frequency of SLO violations, all of which are resolved within a short time window.

\end{abstract}
\section{Introduction}
\label{intro}
Microservice architecture has emerged as a dominant architectural style due to its scalability and flexibility~\cite{Kratzke:17,microservice}. A single system may include tens or hundreds of containerized microservices, each of which may need to scale independently under fluctuating workloads. As a result, auto-scaling is both essential and difficult. It must consider microservice interactions and workload heterogeneity. At the same time, it needs to balance the risk of service-level objectives (SLOs) violations caused by insufficient resources and wasteful over-provisioning due to  excessive replica allocation~\cite{Aksakalli:21,Nunes:24}.

Auto-scaling microservice-based systems is therefore a complex task, especially because the goal is to ensure that the system collectively meets its SLOs rather than simply reacting to individual resource metrics such as CPU or memory usage~\cite{Ding:19}. Improper scaling can lead to SLO violations, causing degraded service quality and potential loss of user satisfaction, or to over-provisioning, which in turn increases operational cost. Since many service providers, such as online banking platforms, e-commerce businesses, and entertainment services, rely on cloud providers for computing power, storage, and GPUs, this situation creates a  \hbox{\emph{fundamental trade-off}}:  provisioning too few resources leads to under-scaling and SLO violations, while provisioning too many resources results in unnecessary costs. Achieving the \emph{``Goldilocks'' zone} -- where resource allocation is neither too little nor too much -- is therefore critical for balancing service quality and operational cost.

Several auto-scaling approaches, e.g.,~\cite{Balla:20,Pramesti:22}, rely on  metrics tied to individual microservices, such as the response time of a single service, which are insufficient to ensure system-level SLOs. This limitation also appears in Kubernetes' default auto-scaling, which acts based on CPU and memory utilization~\cite{kubernetes,Deng:24,Nguyen:20}, and has similarly been shown to fall short in maintaining  SLOs~\cite{Nunes:24}. While some SLO-aware auto-scaling approaches exist, they typically focus on a specific type of SLO, e.g., latency,  and only react once violations occur~\cite{Qiu:20}, offering no safeguards against over-provisioning, i.e., overreacting, when fewer resources would suffice for ensuring SLOs. Approaches that seek to address both over- and under-provisioning generally remain reactive, repeatedly issuing one-off scaling actions without exploiting knowledge from past adaptations~\cite{Nunes:24}, and therefore cannot improve decisions over time or proactively prevent SLO violations.

In this paper, we propose \framework, an auto-scaling framework for  microservices. We take the view that \emph{effective auto-scaling requires learning better scaling behaviour over time, rather than repeatedly generating isolated adjustments at runtime.}  Accordingly, \framework\ evolves reusable scaling logic that generalizes across changing workloads and environmental contexts in microservice-based systems. To enable this, \framework\ uses \textit{Genetic Programming (GP)}~\cite{Luke:13} to automatically evolve and refine scaling formulas that compute the number of microservice replicas directly from runtime metrics. These formulas act as adaptive scaling policies that can be incrementally improved and repeatedly applied when metric values change.
Rather than relying on predefined heuristics or fixed threshold rules, GP searches a rich space of scaling strategies and incrementally improves them using runtime feedback 
-- favouring solutions that both reduce resource consumption and lower the frequency and severity of SLO violations. To enable GP to run efficiently online, \framework\ first constructs two prerequisite artifacts offline: (1) a list of bottleneck microservices whose replica counts will be dynamically determined, and (2) a surrogate model trained to predict SLO outcomes for candidate scaling strategies. During adaptation, GP targets these bottleneck microservices and generates candidate formulas to determine their replica counts, evaluates each candidate using the surrogate model, and selects the most effective strategy, which is then applied to the running system.

We evaluate \framework\ on two systems -- a CPU-intensive online shopping system, and a GPU-based chatbot system built on  Large Language Models (LLMs). Compared to the Kubernetes Horizontal Pod Autoscaler (HPA), \framework\ reduces computational resource usage by 50.65\% for the shopping system and 46.4\% for the chatbot. Compared to its random-search variant, \framework\ achieves a statistically significant reduction in pod usage for the chatbot system, and consistently reduces average SLO violations across both case studies.

The main \textbf{novelty} of our work is the systematic exploration and evaluation of the behaviour of microservice-based systems under different scaling strategies, with a focus on search-based software engineering (SBSE) techniques, specifically GP, with random search serving as a baseline. The \textbf{significance} of our work is the application of SBSE to an important and rapidly emerging problem: automated scaling of complex systems that are increasingly composed of LLM-based components and deployed on microservice architectures. %

\section{Motivation}
\label{motivation}
We motivate our work through an SLO-driven auto-scaling scenario based on our case-study systems. SLOs capture core quality-of-service requirements and indicate the contractual expectations established with clients. Our case studies define two SLOs:

\noindent(\textbf{SLO1})~90\% of the requests shall be responded to within 500ms.

\noindent(\textbf{SLO2})~The success rate of the requests shall be more than 95\%.

SLO1 tracks the 90th-percentile latency; exceeding 500\,ms signals an SLO violation (under-provisioning), while consistently low latency indicates over-provisioning. SLO2 measures reliability; a success rate below 95\% indicates a violation, while significantly higher rates imply that the system is provisioned with too many resources.

\framework{} dynamically adjusts microservice replicas to fix SLO violations and avoid over-provisioning. It periodically checks SLOs, collects contextual metrics such as incoming query load (measured in queries per second, QPS), CPU usage, memory consumption, and network traffic, and uses GP to generate scaling policies that scale replicas up or down accordingly.

\section{Our Approach (\framework)}
\label{approach}

Figure \ref{fig3:controlloop} shows the self-adaptation control loop of \framework, which dynamically adjusts microservice replica counts to prevent both under-provisioning (SLO violations) and over-provisioning (unnecessary resource usage). \framework\ takes as input a single SLO, specified by one target metric and a fixed threshold for that metric. For example, SLO1 (Section \ref{motivation}) uses the 90th-percentile request latency as the SLO metric with a threshold of 500ms, while SLO2 uses the request success rate metric with a 95\% threshold.  Before \framework's monitoring starts, it first builds two prerequisite artifacts offline, as described below:

\begin{figure}[t]
    \centering
    \includegraphics[width=0.85\linewidth]{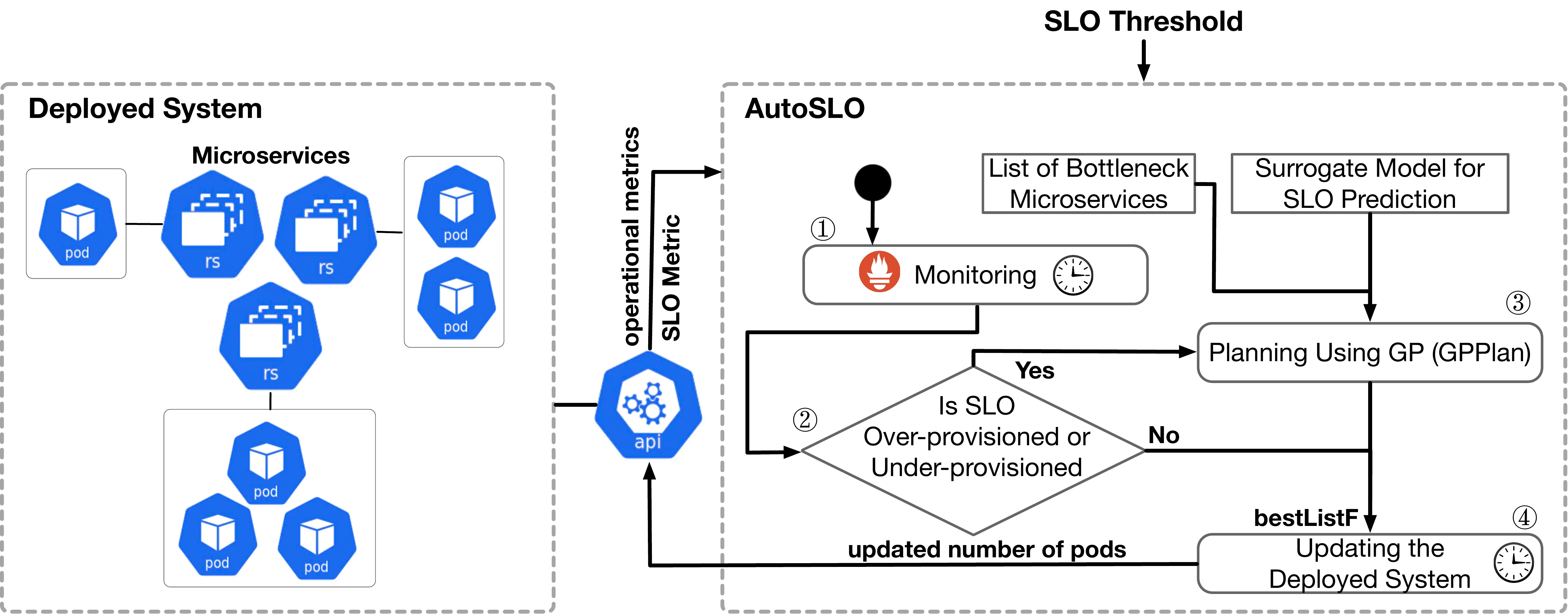}
         \vspace*{-.2cm}
    \caption{Self-adaptation control loop of \framework.}
\label{fig3:controlloop}
     \vspace*{-.2cm}
\end{figure}

\textbf{Prerequisite~1: List of Bottleneck Microservices.} In microservice-based systems, a small subset of services can become overloaded under high workloads and degrade end-to-end performance; we refer to these as \emph{bottleneck microservices}~\cite{Xie:24}. Scaling non-bottleneck services typically has limited effect on SLO satisfaction. Therefore, \framework\ focuses on scaling only bottleneck microservices to reduce the search space and speed up the discovery of effective scaling strategies. To identify bottlenecks, we adopt Performance Bottleneck Analysis (PBA)~\cite{Xie:24}, which detects performance degradation and resource wastage under varying workloads. Accordingly, \framework\ adjusts the replica counts of the identified bottleneck microservices to address both under- and over-provisioning.

\textbf{Prerequisite~2: A Surrogate Model for SLO Prediction.} Evaluating each candidate scaling strategy by deploying it online is too expensive. Instead, \framework\ trains a regression-based surrogate model~\cite{NejatiSSFMM23,LiuNLB19} offline and uses it during planning to evaluate the candidate scaling strategies generated by GP. For each training record, we log the operational metrics including CPU usage, memory consumption, and queries per second (QPS), the pod counts of the bottleneck microservices, and the observed SLO metric value. During GP planning, each candidate strategy is first evaluated to obtain its recommended pod counts (bounded by the allowable minimum and maximum), and the surrogate then predicts the resulting SLO metric value (e.g., 90th-percentile request latency for SLO1 or request success rate for SLO2).

With the two prerequisites artifacts in place, \framework\ operates as a closed-loop controller that continuously monitors the deployed system and updates microservice replica counts through a four-step adaptation loop (Figure~\ref{fig3:controlloop}). \framework\ targets the common \emph{one-container-per-pod} deployment model in Kubernetes~\cite{Schmidt:23}, where each microservice replica corresponds to a single pod; thus, replica count equals pod count. We therefore use \emph{pods} and \emph{replicas} interchangeably in the remainder of this paper. We now describe the four steps of the \framework\ loop:

\textbf{Step 1. Monitoring:} \framework\ periodically retrieves operational metrics and SLO metrics from the deployed system using Prometheus~\cite{Prometheus}, an open-source microservices monitoring toolkit. Metrics exposed by Kubernetes components or built-in exporters, including CPU usage and memory consumption, are directly collected by Prometheus, whereas workload-related operational metrics (e.g., QPS)  and SLO metrics (e.g., 90th-percentile request latency for SLO1 or request success rate for SLO2) are obtained with the support of Istio~\cite{istio}. These metrics serve as inputs to Step 3 for determining the recommended number of pods for the  bottleneck microservices.

\textbf{Step 2. SLO checking:}  \framework\ evaluates current SLO metrics against their thresholds to determine whether an SLO is violated (under-provisioned) or exceeded (over-provisioned). For example, SLO1 and SLO2 (Section~\ref{motivation}) are violated when the 90th-percentile response time exceeds 500 ms and when the request success rate drops below 95\%. The system is considered over-provisioned when the response time is, for instance, below 300 ms and the success rate reaches 98\%, indicating unnecessary resource allocation. The acceptable margin for identifying over-provisioning is user-configurable. If the SLO is either violated or exceeded, \framework{} proceeds to planning (Step 3); otherwise, it continues updating the deployed system (Step 4) based on prior scaling decisions.

\textbf{Step 3. Planning using GP:} In this step, \framework{} invokes the \emph{GPPlan procedure} (Algorithm~\ref{algo3:gp}) to generate an updated scaling strategy. GPPlan produces \textit{bestListF}, a list of learned formulas -- one formula for each bottleneck microservice -- that specifies how pod counts should be computed from the operational metrics. Each element in \textit{bestListF} is a mathematical expression (e.g., a combination of operational metrics and constants) that, when evaluated, computes the recommended number of pods for its corresponding bottleneck microservice. To ensure that the knowledge gained in earlier planning cycles is preserved, GPPlan always includes the best solution from the previous invocation -- stored as \textit{BestSol}  -- in its initial population (line 2, Algorithm~\ref{algo3:gp}).

\setcounter{AlgoLine}{0}
\begin{algorithm}[!t]
\scriptsize
\DontPrintSemicolon
\SetSideCommentRight
\SetNoFillComment 
\KwIn{\emph{SLOThreshold}\;}
\KwIn{\emph{$podNum_{max}$/$podNum_{min}$}: Maximum/Minimum allowable pod number\;}
\KwIn{\emph{operationalMetrics}\;    }
\KwIn{\emph{BestSol}:  Best solution from the previous round\;}
\SetKwInOut{Output}{Output}
\Output{bestListF}
     $t=0$\;
     $P_0 = \{\mathit{BestSol}\}$\;
    \While { not (stopCondition)} 
    {
    \uIf{t==0}{
    \textit{offsprings}=InitialPopulation()\;
    }
    \Else{
    \textit{offsprings}=Breed$(P_t)$\;
    }
    \For{$w \in$ offsprings }{
    $\widehat{SLO}_w$ =Surrogate($w, operationalMetrics, podNum_{max}, podNum_{min})$\;
    $w.fit$=Fitness($w, \textit{SLOThreshold},  \widehat{SLO}_w$)\;
    }
    $P_{t+1}$= Select($\textit{offsprings}\cup P_t$) \;
    t=t+1\;
    }
    bestListF=BestIndividual($P_t$)\;
    \caption{GPPlan used in Step~3 of \framework{} (see Figure~\ref{fig3:controlloop}).}
    \label{algo3:gp}
\end{algorithm}

GPPlan begins by retrieving the operational metrics from the monitoring step and constructing an initial population of candidate individuals (line 5, Algorithm~\ref{algo3:gp}). Each individual is a list of formulas that compute the number of pods for each bottleneck microservice, where each formula is a mathematical expression defined by the grammar below, specifying how operators, constants, and metrics can be combined to form valid scaling expressions.

        \fbox{\parbox{\dimexpr\linewidth-9\fboxsep-2\fboxrule\relax}{\centering exp ::= 
        exp\,$+$\,exp\,$|$
        exp\,$-$\,exp\,$|$
        exp\,$*$\,exp\,$|$
        exp\,$/$\,exp\,$|$
        const\,$|$ metrics}}

In the above grammar, ``$\mid$'' separates alternatives; \textit{const} denotes an ephemeral random constant generator~\cite{Veenhuis13} that produces integers between 1 and 100; and \textit{metrics} refers to the operational metrics collected during monitoring.

\begin{figure}[t]
    \centering
    \includegraphics[width=0.55\linewidth]{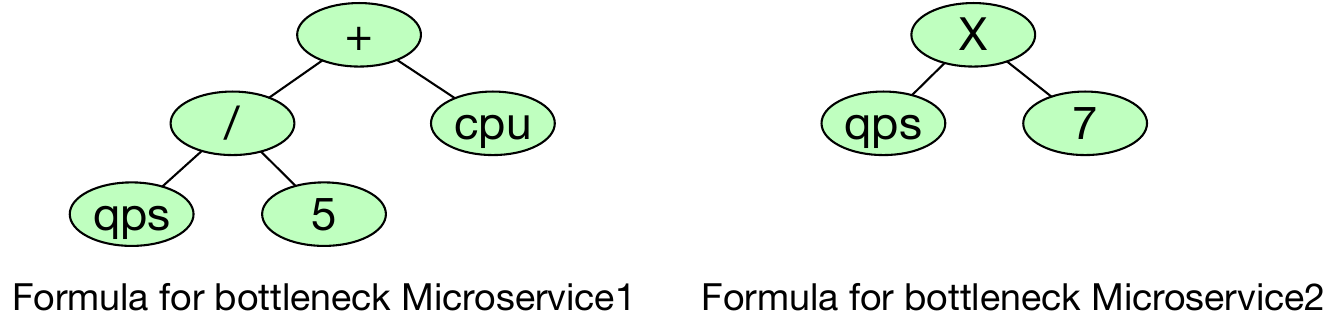}
     \vspace*{-.2cm}
    \caption{A  GP individual with formula trees for two bottleneck microservices.}
    \label{fig:parseTree}
    \vspace*{-.1cm}
\end{figure}

To illustrate, consider a system with two bottleneck microservices, \textit{microservice1} and \textit{microservice2}. A GP individual for this system is a two-element list, where each element is a formula --~generated using the grammar described above~-- and represented as a parse tree. Figure~\ref{fig:parseTree} shows an example GP individual in which the formula ``$\mathit{qps/5 + cpu}$'' determines the number of pods for the first bottleneck microservice, and ``$\mathit{qps \times 7}$'' determines the number for the second, where \textit{qps} denotes the queries-per-second operational metric, and \textit{cpu} represents the CPU consumption of the microservice.

GPPlan generates its initial population by randomly constructing  parse trees such as those in Figure~\ref{fig:parseTree} using the grow method~\cite{poli2008field}. In each tree, the root and internal nodes are labelled with mathematical operators, while the leaf nodes are labelled with either constants or  operational metrics.

For each generated individual $w$, GPPlan invokes the surrogate model (pre-computed offline; Prerequisite~2) to estimate the predicted SLO value, $\mathit{\widehat{SLO}}$ (line~9, Algorithm~\ref{algo3:gp}). The formulas in $w$ are evaluated to compute the recommended pod counts for each bottleneck microservice, which -- along with the operational metrics and pod count bounds -- are supplied to the surrogate model to obtain $\mathit{\widehat{SLO}_w}$. \framework{} then computes the fitness of $w$ based on this predicted outcome (line~10, Algorithm~\ref{algo3:gp}).

The fitness  for each individual measures whether the corresponding scaling strategy violates the SLO and, if so, by what extent. If the strategy satisfies the SLO, the fitness instead reflects resource efficiency, i.e., the extent to which it minimizes pod usage. An individual represents a desirable scaling strategy when it satisfies the SLO while also minimizing computational resource consumption.  More precisely, for each individual $w$, if $\mathit{\widehat{SLO}_w}$ exceeds \textit{SLOThreshold}, then the strategy is predicted to violate the SLO, and its fitness is computed as:

\begin{center}
$\begin{array}{c}
\frac{\lvert SLOThreshold - \widehat{SLO}_w \rvert}{SLOThreshold} + 1
\end{array}$
\end{center}

The above expression quantifies the severity of deviation from the SLO threshold, and the additive constant ensures that all violating individuals receive a fitness value strictly greater than one. This allows SLO-violating strategies to be clearly distinguished from SLO-satisfying ones. Otherwise, when the individual satisfies the SLO, i.e., when $\mathit{\widehat{SLO}_w}$ is less than or equal to \textit{SLOThreshold}, the fitness is computed only in terms of resource efficiency:
$\frac{\mathit{totalPod_w}}{\mathit{podNum}_{\text{max}} \times N}$
where $\mathit{totalPod}_w$ is the total number of pods of bottleneck microservices required by individual $w$, $\mathit{podNum}_{\text{max}}$ is the maximum allowable pod count per microservice, and $N$ is the number of bottleneck microservices.

GPPlan is a minimization optimization algorithm that favours individuals with the lowest fitness values, representing scaling strategies that satisfy the SLO using fewer pods. It evolves the population using standard GP operators: one-point crossover and one-point mutation for breeding (line~7, Algorithm~\ref{algo3:gp}), and tournament selection for the next generation (line~11). The process continues until the stopping criterion, a maximum number of iterations, is reached, after which GPPlan returns the optimal individual, \textit{bestListF} (line~13).

\textbf{Step 4. Updating the Deployed System:} In this step, \framework{} applies the auto-scaling strategy described by \textit{bestListF}. As shown in Figure~\ref{fig3:controlloop}, if Step~2 detects SLO over- or under-provisioning, \framework{} first re-invokes GPPlan to compute an updated \textit{bestListF} and then Step~4 is invoked. Otherwise, Step~4 is executed after the monitoring step and uses the most recently learned \textit{bestListF}. 

\framework{} evaluates each formula in \textit{bestListF} using the latest operational metrics collected from the deployed system. Each formula outputs a recommended pod count for its corresponding bottleneck microservice. If this recommended count differs from the pod count currently deployed, \framework{} issues a scaling request via the Kubernetes API to adjust pod numbers accordingly. Because the one-container-per-pod model is commonly adopted in Kubernetes deployments~\cite{Schmidt:23}, \framework{} interprets pod adjustments as direct changes to microservice replica counts. Scaling operations occur asynchronously, while \framework{} continues monitoring until the next control-loop cycle is triggered. This enables responsive scaling while avoiding unnecessary GP invocations when existing strategies remain valid.

\section{Evaluation}
\label{evaluation}
We evaluate \framework\ by addressing the following research question (RQ):

\noindent\textbf{RQ.} \textit{How effective is \framework\ at scaling microservice-based systems under dynamic workloads to meet their SLOs without over-provisioning computational resources?}
To answer this question, we evaluate \framework\ on two case-study systems with distinct resource profiles: a CPU-intensive application and a GPU-intensive LLM-based application. CPU-intensive services allow elastic, fine-grained resource allocation, whereas GPU-intensive services are constrained by coarse-grained and scarce resources. The selection of these two systems enables us to assess whether \framework\ generalizes across both microservice environments. %

\noindent\textbf{Case Studies.} For the CPU-intensive scenario, we use Boutique Shop~\cite{boutique}, an open-source microservice-based e-commerce application developed by Google and widely used as a benchmark system. It consists of eleven microservices supporting common user interactions such as browsing products, managing carts, and completing purchases.

For the GPU-intensive scenario, we evaluate an open-source LLM-based chatbot from the existing software engineering literature~\cite{DBLP:conf/icsm/ChaudharyVTNS24}, developed using LLaMA-3~\cite{llama3}, which we refer to as Chatbot. Our evaluation focuses on the scalability and resource efficiency of handling query workloads, not on the correctness of the generated answers by Chatbot. To bound the scalability of Chatbot within the physical constraints of the GPU hardware used in our experiments, we employ GPU time slicing on an NVIDIA A40 GPU and cap the amount of GPU memory allocated to each pod. Based on exploratory experiments on this setup, the cluster can support a maximum of three pods simultaneously.

We evaluate the Boutique Shop case study based on SLO1 from in Section~\ref{motivation}, i.e., 90th-percentile request latency $\mathit{\leq}$ 500ms.  Boutique Shop serves user-facing e-commerce requests where response time directly affects user experience; as such, we follow prior work and evaluate it using SLO1 which is a latency-based SLO~\cite{Xie:24}. SLO1, however,  is not suitable for the Chatbot case study because latency-based SLOs do not reflect the dominant failure modes of chatbot applications. As Chatbot reaches service capacity, performance degradation appears as request failures (HTTP 500 errors) rather than increased latency, rendering latency measurements unstable and unreliable. Hence, for the Chatbot case study, we use SLO2 from Section~\ref{motivation} (request success rate $\mathit{\geq}$ 95\%).

\noindent\textbf{Implementation.} Our case-study systems run on a Kubernetes cluster consisting of one master and two worker nodes. All nodes run Ubuntu and have two Intel Xeon Gold 6338 CPUs, 512GB of memory and one A40 GPU.  \framework\ is implemented in Python and builds on open-source  PBScaler~\cite{Xie:24}, to identify bottleneck microservices. We collect runtime metrics via Prometheus~\cite{Prometheus} integrated with Istio~\cite{istio}, generate traffic using IoTECS~\cite{Li:24}, and implement GPPlan using DEAP’s genetic programming module (v1.0)~\cite{deap}.

\noindent\textbf{Baselines.} We use two baselines. Our first baseline is the Horizontal Pod Autoscaler (HPA), the industry-standard auto-scaling mechanism provided by Kubernetes. HPA adjusts the number of pods for each microservice based on resource utilization metrics such as CPU or memory usage, which represent a subset of operational metrics. Following Nguyen et al.~\cite{Nguyen:20}, which is widely used as a reference configuration in prior auto-scaling studies, we employ CPU utilization with a fixed target threshold and the default, unmodified Kubernetes controller. This setup represents the canonical HPA configuration adopted both in practice and in the research literature~\cite{Nguyen:20}. Unlike \framework, which operates at the application level and bases decisions on SLO satisfaction, HPA performs reactive, per-microservice scaling driven exclusively  by resource utilization signals. The HPA CPU utilization threshold is set to 80\%, a value commonly adopted in prior studies~\cite{Xie:24}. This setting balances responsiveness and stability: it allows the system to scale up when CPU usage becomes significant, while avoiding frequent, unnecessary scaling events during minor workload fluctuations.

Our second baseline is a random-search variant of \framework, denoted as \framework-Ran. 
In \framework-Ran, the monitoring and analyzing steps are identical to those in \framework; however, the GP-based planning procedure (GPPlan) is replaced with a random-search strategy. Instead of evolving scaling formulas, \framework-Ran directly generates candidate pod counts for bottleneck microservices and evaluates them using the same fitness function as \framework. Because scaling decisions are produced as concrete pod values rather than formulas, \framework-Ran cannot proactively adapt to changes in workload or operational metrics. As a result, it must re-enter the planning phase whenever an SLO violation or exceedance occurs and lacks a mechanism for anticipating future scaling needs. This baseline, in addition to being a standard in SBSE  as it uses random search,   allows us to isolate the contribution of GP-based planning in proactively reducing SLO violations and over-provisioning under dynamic workloads.

\noindent\textbf{Experiments.} We simulate user inputs using a dynamic workload that alternates between normal-load and high-load phases. During normal-load phases, the request rate remains low enough for the systems to meet their SLOs, while high-load phases increase the request rate to trigger SLO violations. Each experiment runs for a fixed duration and includes multiple escalation and de-escalation cycles, allowing us to evaluate scaling behaviour under dynamic conditions.

We configure the two prerequisites of \framework\ (see Figure~\ref{fig3:controlloop}), i.e., a list of bottleneck microservices and a surrogate model for SLO prediction, for both case-study systems: (1)~We identify the bottleneck microservices using PBScaler~\cite{Xie:24}, which reports the \textit{frontend} and \textit{productcatalogservice} microservices as bottlenecks in the Boutique Shop system, and the microservice responsible for serving LLM inference as the bottleneck in the Chatbot system. (2)~To construct a surrogate model for each case study, we collect training data over a 10-hour period by generating randomized workload patterns (using IoTECS) while varying the number of pods for each bottleneck microservice. Using this dataset, we train and evaluate multiple ML regression models and select the Random Forest Regressor~\cite{breiman:11} as the best-performing surrogate model, i.e., the one with the highest $R^2$ score and lowest mean absolute error (MAE), for SLO prediction in both systems. The training data and experiment results, i.e., $R^2$ scores and MAE values,  for building a surrogate model are available online~\cite{AutoSLO}.

Table~\ref{tab:parameters} shows the configuration  parameters used for \framework. For the mutation and crossover rates, the maximum tree depth and the tournament size, we chose recommendations from either the GP literature~\cite{poli2008field,LukeP06} or DEAP's documentation~\cite{deap}. We set the update duration (Step~4 in Figure \ref{fig3:controlloop}) to 15s, which corresponds to the time required for Kubernetes to apply updates and for Prometheus \cite{Prometheus} to provide refreshed metrics.

\begin{table}[t]
\caption{Parameters of \framework\ for our experiments.}
\label{tab:parameters}
\vspace*{-1em}
\begin{center}
\scalebox{.78}{
\begin{tabular}{p{4.5cm} p{0.6cm}||p{9cm} p{1cm}}
\toprule
Mutation rate  &  0.1 & \# of Generations (Botique Shop/Chatbot)  &  30/20 \\
\hline
Crossover rate &  0.9 & Population size &  50  \\
\hline
Maximum depth of GP tree &  15 & Frequency of updating the deployed system & $15$s\\
\hline
Tournament size &  3  &  \framework\ control-loop cycle duration (GPPlan trigger period)  &  $1$min  
\\
\bottomrule
\end{tabular}}
\vspace*{-.3cm}
\end{center}
\end{table}

To choose the appropriate population size and number of generations for GPPlan, we ensure that the total time required to plan and apply a scaling action fits within a single cycle of the \framework\ control loop. In other words, \framework\ must complete its planning and execute any scaling action before the next monitoring cycle begins. The total processing time includes the GP computation as well as Kubernetes-related delays, such as API calls and the time needed to create, initialize, or remove pods. In our work, this total processing time is 1min, as indicated in Table~\ref{tab:parameters}.  Guided by this requirement and preliminary experiments on our case-study systems, we configure the GP parameters as follows. For the  Boutique Shop system, GPPlan uses a population size of 50 and runs for 30 generations, which keeps the combined planning and execution time within the one-minute control-loop cycle. For the Chatbot system, pod initialization is slower, so while the population size remains 50, we reduce the number of generations to 20 to ensure GPPlan also completes within the same one-minute window. For a fair comparison, \framework-Ran is configured using the same population size and number of generations as GPPlan on each system.

For the  Boutique Shop system, we configure the allowable number of pods between one and ten, providing adequate elasticity within the experimental Kubernetes cluster to accommodate workload variations. In contrast, for the Chatbot system, the number of pods per microservice ranges from one to three, reflecting the maximum pod capacity supported by the underlying hardware. At the start of each experiment, one pod is deployed for every microservice. We repeated each experiment ten times to account for random variation.

\begin{figure}[htbp!]
    \centering
    \includegraphics[width=0.8\linewidth]{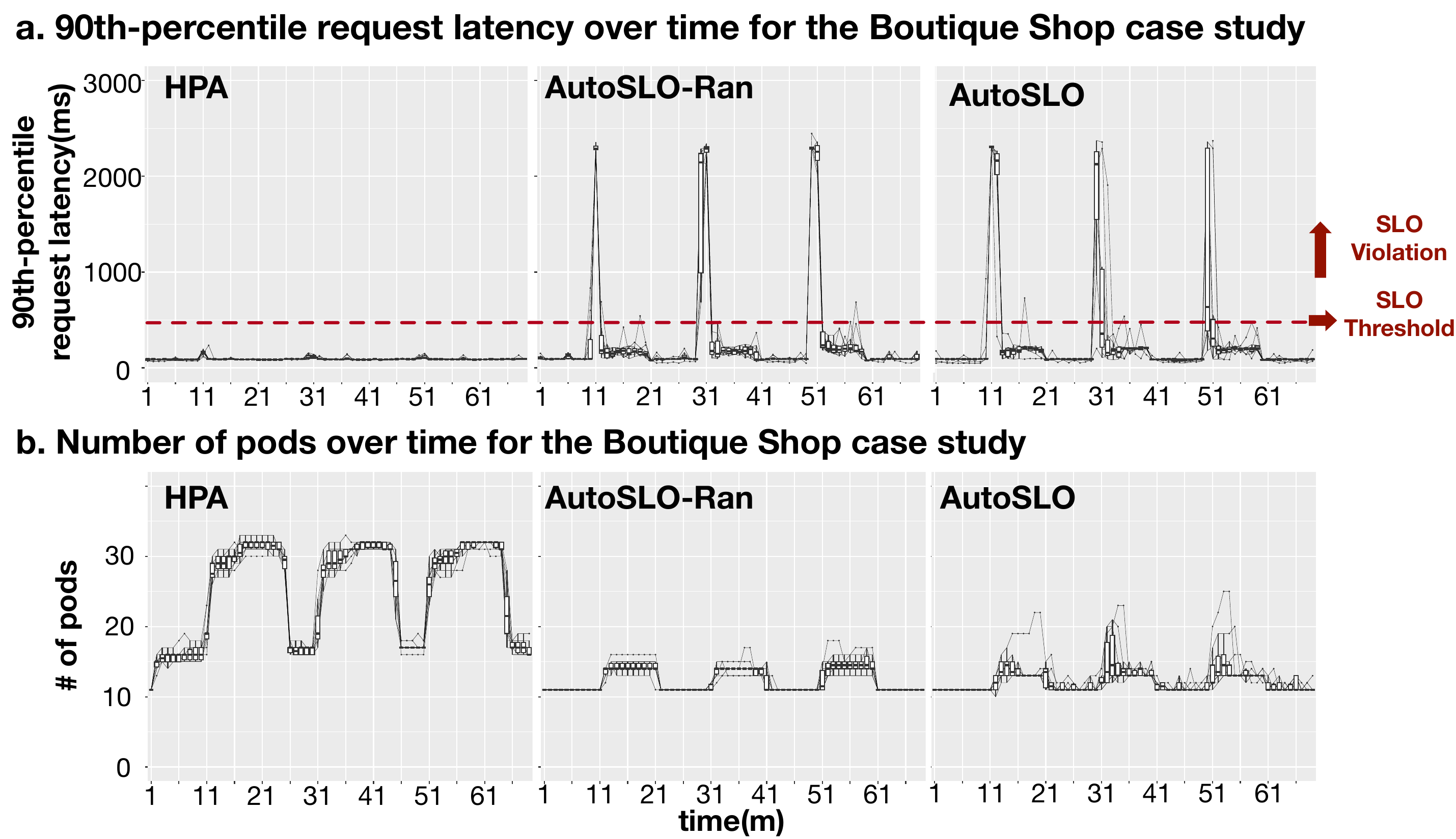}
    \caption{Experimental results for  the Boutique Shop (CPU-intensive) system: (a) 90th-percentile request latency (SLO1) and (b) number of pods  across ten repeated applications of a dynamic workload. The results compare \framework, \framework-Ran, and HPA, showing how each approach responds to workload changes and scales pods while impacting SLO compliance.}    \label{fig3:boutiqueData}
    \vspace*{.3cm}
    \includegraphics[width=0.8\linewidth]{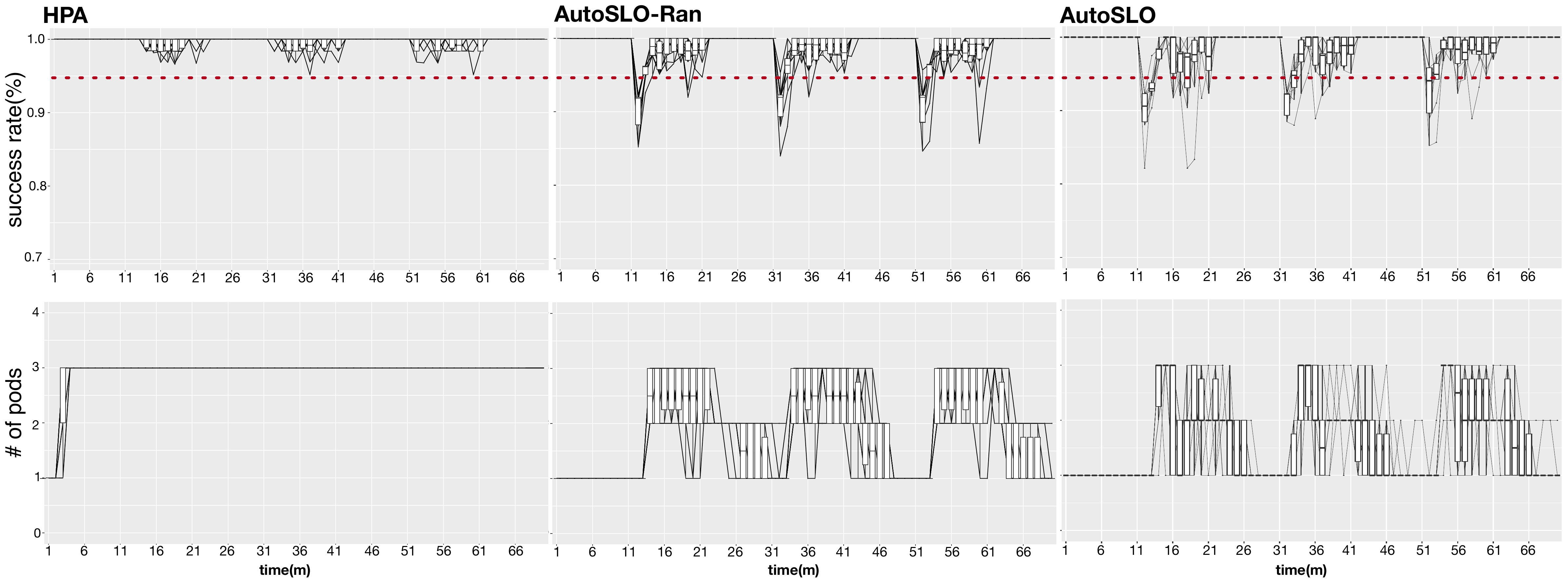}
\caption{Experimental results for  the Chatbot (GPU-intensive) system:(a) Request success rate (SLO2) and (b) number of pods across ten repeated applications of a dynamic workload. The results compare \framework, \framework-Ran, and HPA, illustrating differences in SLO performance and scaling behaviour under repeated high-load conditions.}
    \label{fig3:chatbotData}
    \vspace*{-0.5cm}
\end{figure}

\noindent\textbf{Results.} Figures~\ref{fig3:boutiqueData} and \ref{fig3:chatbotData} present the experimental results obtained by applying \framework, \framework-Ran, and HPA to the  Boutique Shop and Chatbot systems, respectively. Specifically, Figure~\ref{fig3:boutiqueData}(a) illustrates the variations in the SLO1 metric -- representing 90th-percentile request latency -- over time for the  Boutique Shop case study under a dynamic workload. Similarly, Figure~\ref{fig3:chatbotData}(a) shows the request success rate (SLO2) over time for the Chatbot system. For both systems, the workload was applied for approximately one hour, during which the SLO1 metric increased and the SLO2 metric decreased during periods of high input workload. Dashed red lines in both figures indicate the thresholds for SLO violations, where any instance of the data curve crossing these lines represents a violation. Figures~\ref{fig3:boutiqueData}(b) and \ref{fig3:chatbotData}(b) display the number of pods required by each scaling approach in response to these workloads. 

 As the figures show, under HPA, no SLO violations are observed. This is, however, because HPA consistently requires a higher number of pods than \framework\ and \framework-Ran to avoid SLO violations. In contrast, both \framework\ and \framework-Ran experience SLO violations during periods of high workload. Nevertheless, each violation is resolved within at most two control-loop cycles ($<2$ minutes) by increasing the number of pods allocated to the bottleneck microservices (see Figure~\ref{fig3:boutiqueData}(b)). The 2-minute bound is primarily determined by our 1-minute control-loop period and Kubernetes pod rollout latency, and can be reduced by shortening the control-loop interval and using faster pod provisioning.  Both \framework\ and \framework-Ran are able to reduce the number of pods when  the workload transitions from high to low.

To compare \framework\ with the two baselines, we assess the average number of violations and the average number of  required pods across ten runs using the Wilcoxon Rank-Sum test~\cite{wilcoxon1992individual} and the Vargha–Delaney effect size ($A_{12}$)~\cite{vargha2000critique}, adopting a 1\% significance level. For both metrics, smaller values indicate better performance. A difference is deemed statistically significant when the \hbox{$p$-value~$< 0.01$}. Effect sizes are interpreted as small, medium, or large when $A_{12}$ deviates from 0.5 by at least 0.06, 0.14, and 0.21, respectively. Table~\ref{tab:autoslo-comprehensive} shows the statistical test results for the number of pods and number of violations.

\begin{table}[t]
\centering
\caption{Averages and statistical-test results comparing \framework, \framework-Ran (Ran) and HPA with respect to  the average number of violation occurrences and average number of required pods by 10 runs of \framework\ versus baselines for  the  Boutique Shop and Chatbot systems. The effect sizes are labelled as Large (L), Medium (M), Small (S) and Negligible (N). The $p$-values highlighted in blue represent cases where \framework\ significantly outperforms HPA or \framework-Ran.  \framework\ is not outperformed by \framework-Ran in any comparison. 
Although \framework\ is outperformed by HPA in terms of the number of violations, HPA uses twice as many pods (resources) than \framework.}
\label{tab:autoslo-comprehensive}
\resizebox{\textwidth}{!}{
\begin{tabular}{|l|l|c|c|c|cc|cc|}
\hline
\multirow{2}{*}{\textbf{Case Study}} & \multirow{2}{*}{\textbf{Metric}} 
& \textbf{\framework} & \textbf{HPA} & \textbf{Ran} 
& \multicolumn{2}{c|}{\textbf{\framework\ vs. HPA}} 
& \multicolumn{2}{c|}{\textbf{\framework\ vs. Ran}} \\ \cline{3-9}
 &  & \textbf{Avg.} & \textbf{Avg.} & \textbf{Avg.} 
 & \textbf{$p$-value} & \textbf{Effect ($A_{12}$)} 
 & \textbf{$p$-value} & \textbf{Effect ($A_{12}$)} \\ \hline

\multirow{2}{*}{\begin{tabular}[c]{@{}l@{}}\textbf{ Boutique Shop}\\(CPU-Intensive)\end{tabular}}
 & \# of Pods & 12.36 & 25.05 & 12.35 
 & {\cellcolor{cyan!20}$2.2 \times 10^{-16}$} & $0.03$ (L) 
 & $0.76$ & $0.50$ (N) \\ \cline{2-9}
 & \# of Violations & 2.6 & 0 & 3.0 
 & $4.13 \times 10^{-5}$ & $1.00$ (L) 
 & $0.078$ & $0.35$ (S) \\ \hline

\multirow{2}{*}{\begin{tabular}[c]{@{}l@{}}\textbf{Chatbot}\\(GPU-Intensive)\end{tabular}}
 & \# of Pods & 1.57 & 2.93 & 1.79 
 & {\cellcolor{cyan!20}$<2.2 \times 10^{-16}$} & $0.08$ (L) 
 & {\cellcolor{cyan!20}$9.02 \times 10^{-6}$} & $0.44$ (N) \\ \cline{2-9}
 & \# of Violations & 2.8 & 0 & 3.0 
 & $3.29 \times 10^{-5}$ & $1.00$ (L) 
 & $0.168$ & $0.40$ (S) \\ \hline
\end{tabular}
}

\end{table}

As shown in Table~\ref{tab:autoslo-comprehensive}, \framework\ significantly outperforms HPA across both case studies in terms of resource efficiency, achieving statistically significant reductions in average pod usage while entailing fewer than 2.8 violations on average that are all resolved in less than 2 min. For the  Boutique Shop system, \framework\ requires an average of 12.36 pods compared to HPA’s 25.05, a 50.65\% reduction in average pod usage. Similarly, in the GPU-intensive Chatbot system, \framework\ lowers average pod usage from 2.93 (HPA) to 1.57, a 46.4\% reduction.  Against the \framework-Ran baseline, \framework\ performs on par in pod reduction for the CPU-intensive system, while in the Chatbot case, it yields a statistically significant improvement, reducing average pod usage from 1.79 to 1.57. \framework\ also incurs fewer average SLO violations relative to \framework-Ran, decreasing violation counts from 3 to 2.6 on the  Boutique Shop system and from 3 to 2.8 on the Chatbot system, though these differences are not statistically significant.

\vspace*{.2cm}
\resq{The answer to RQ: Compared to HPA, \framework\ lowers average pod usage by 50.65\% for the CPU-intensive case-study system and 46.4\% for the GPU-intensive system. Compared to its random-search variant, \framework\ achieves a statistically significant reduction in pod usage for  the GPU-intensive case study, and delivering a consistent reduction in average SLO violations across both case studies. The SLO violations observed for \framework\ reflect a deliberate design trade-off rather than a shortcoming. Unlike HPA, which avoids violations through conservative over-provisioning, \framework\ prioritizes resource efficiency while allowing brief, bounded violations that are automatically resolved through adaptation. In all experiments, violations are  corrected within at most two control-loop cycles (less than two minutes).}

\noindent\textbf{Limitations.} \framework\ assumes a fixed set of bottleneck microservices, whereas real systems may experience shifting bottlenecks that \framework\ cannot currently handle. While \framework\ is currently designed to assess a single SLO, it can be extended to multi-SLO settings by using multiple surrogate models --~one per SLO~-- and by modifying the first component of the fitness function discussed in Step 3 so that it checks for violations of multiple SLOs for each individual. This direction nonetheless requires  experiments to validate effectiveness and efficiency.

\noindent\textbf{Internal validity.} \framework’s effectiveness depends on the accuracy of its surrogate model; mispredictions may lead to rejecting viable scaling strategies or failing to prevent violations. To mitigate this threat, we trained the model using 10 hours of workload data and selected the Random Forest Regressor after comparing multiple regressor models based on the $R^2$ and MAE metrics. Another threat relates to randomness within GP evolution. All experiments were repeated ten times under identical environmental settings to account for randomness. 

A potential threat to internal validity is the runtime overhead of \framework. However, the overhead is minimal: adaptations target only bottleneck microservices, GPPlan runs only when needed, and each adaptation loop completes within roughly one cycle, causing no significant performance impact.

\noindent\textbf{External validity.} We evaluate \framework\ on two representative microservice-based systems: a CPU-intensive web application and a GPU-intensive chatbot, covering both traditional and emerging workloads. Evaluating \framework\ on additional applications and domains would further substantiate generalizability.
\section{Related Work}
\label{relatedwork}

Auto-scaling for microservice-based systems has been widely studied~\cite{Deng:24,Singh:19,Magdelaine:20,Oyeniran:24}, with existing approaches relying on rule-based mechanisms, machine learning (ML), or search-based optimization. Pozdniakova et al.~\cite{Pozdniakova:18} present a high-level auto-scaling architecture focusing on system layers rather than SLO enforcement. MS-RA~\cite{Nunes:24} combines horizontal and vertical scaling using threshold-based rules informed by SLOs and resource metrics, while Pramesti et al.~\cite{Pramesti:22} propose a rule-based approach guided by ML-based response-time prediction. These methods typically operate at the level of individual microservices and do not model system-wide behaviour or inter-service interactions.

Liu et al.~\cite{Liu:22} reallocate soft resources (e.g., threads and database connections) after scaling to improve utilization, but do not use SLOs to drive scaling decisions. Chamulteon~\cite{Bauer:19} combines reactive and proactive logic with conflict resolution, yet remains rule-based and reacts primarily to request arrivals rather than SLOs. PEMA~\cite{Hossen:22} addresses CPU downscaling by initially over-provisioning to satisfy latency SLOs and then iteratively reducing CPU allocations, but considers only response time and requires multiple iterations. HetSev~\cite{Mo:23} proposes a heuristic, cost-aware auto-scaling approach for GPU-based ML inference on Kubernetes that minimizes GPU costs under latency SLOs, but lacks fine-grained microservice scalability. PBScaler \cite{Xie:24} focuses on bottleneck identification and reactive scaling: it detects bottleneck microservices using a random-walk analysis and derives scaling actions from offline logs via genetic algorithms. In contrast, our approach uses bottleneck awareness only to reduce the search space and instead learns reusable scaling formulas online from live system feedback using GP, triggering planning only when SLO violations are imminent. Although both target bottleneck services, they differ fundamentally in adaptation goals (log-driven action selection vs. online policy learning), operational assumptions, and control granularity. While GP has been explored in other self-adaptive settings (e.g., network congestion management \cite{Li:24-2}), to our knowledge it has not been applied to online auto-scaling or systematically  evaluated across both traditional microservices and GPU-intensive systems.

\vspace*{-.2cm}
\section{Conclusion}
\label{conclusion}
\vspace*{-.1cm}
\framework{} evolves scaling formulas with genetic programming so replica counts are driven by runtime metrics rather than fixed thresholds. In our two case studies, \framework{} reduces pod usage substantially while keeping SLO violations low overall. Compared to HPA, AutoSLO cuts resource usage by 50.65\% and 46.4\%, and it also improves over random search by producing reusable scaling logic instead of one-off decisions.

\textbf{Data Availability.} All code, evaluation scripts, and experimental data are available online to enable future use and replication~\cite{AutoSLO}.

\section*{Acknowledgment}
We gratefully acknowledge funding from Mitacs Accelerate and NSERC of Canada under the Discovery program.

\bibliographystyle{splncs04}  %
\bibliography{references}   %

\end{document}